\documentclass[a4paper, 11pt]{article}
\usepackage[T1]{fontenc}
\usepackage{jheppub,graphicx,epsf,amssymb,amsbsy,amsfonts,amssymb,amsmath, empheq,mathrsfs,appendix,soul}
\usepackage{hyperref}

\def\be{\begin{equation}}
\def\ee{\end{equation}}
\def\bea{\begin{eqnarray}}
\def\eea{\end{eqnarray}}

\def\a{\alpha}

\def\d{\delta}

\def\f{\phi}

\def\bg{\bar{g}}
\def\beq{\begin{eqnarray}}\def\eeq{\end{eqnarray}}
\def\ba#1\ea{\begin{align}#1\end{align}}
\def\bg#1\eg{\begin{gather}#1\end{gather}}
\def\bm#1\em{\begin{multline}#1\end{multline}}
\def\bmd#1\emd{\begin{multlined}#1\end{multlined}}

\def\a{\alpha}

\def\d{\delta}
\def\D{\Delta}

\def\G{\Gamma}

\def\pa{\partial}

\def\({\left(}
\def\){\right)}
\def\[{\left[}
\def\]{\right]}

\def\a{\alpha}

\def\d{\delta}

\def\f{\frac}

\def\D{\Delta}

\title{All $S$ invariant gluon OPEs on the celestial sphere}

\author{Shamik Banerjee$^{\,a,b}$, Raju Mandal$^{\,a,b}$, Sagnik Misra$^{\,a,b}$, Sudhakar Panda$^{\,a,c}$, Partha Paul$^{\,d}$}
\affiliation[a]{National Institute of Science Education and Research (NISER), Bhubaneswar 752050, Odisha, India}
\affiliation[b]{Homi Bhabha National Institute, Training School Complex, Anushaktinagar, Mumbai 400094, India}
\emailAdd{banerjeeshamik.phy@gmail.com}
\emailAdd{rajuphys002@gmail.com}
\emailAdd{sagnik.misra@niser.ac.in}
\affiliation[c]{Centre For Cosmology and Science Popularization (CCSP), SGT University, Gurugram, Delhi- NCR, Haryana- 122505, India}
\emailAdd{panda@niser.ac.in}
\affiliation[d]{Centre for High Energy Physics, Indian Institute of Science,
C.V. Raman Avenue, Bangalore 560012, India}
\emailAdd{pl.partha13@gmail.com}

\abstract{$S$ algebra is an infinite dimensional Lie algebra which is known to be the symmetry algebra of some gauge theories. It is a "coloured version" of the $w_{1+\infty}$. In this paper we write down all possible $S$ invariant (celestial) OPEs between two positive helicity outgoing gluons and also find the Knizhnik-Zamolodchikov type null states for these theories. Our analysis hints at the existence of an infinite number of $S$ invariant gauge theories which include the (tree-level) MHV-sector and the self-dual Yang-Mills theory.}

\begin{document}
\maketitle
\flushbottom

\section{Introduction}

S-matrix is an important observable in any quantum field theory in asymptotically flat spacetime. In fact in quantum theory of gravity S-matrix is argued to be the only observable. Therefore any holographic dual theory has to compute the S-matrix elements or scattering amplitudes in the bulk. Celestial holography is an attempt in this direction \cite{Strominger:2017zoo,Pasterski:2021rjz,Donnay:2023mrd}. The realization that the soft theorems of gauge theories and gravity are the Ward identities for different asymptotic symmetries \cite{Strominger:2013lka,Strominger:2013jfa,He,Strominger:2014pwa,Campiglia:2016jdj,Kapec:2016jld, Kapec:2014opa, Banerjee:2020vnt,He:2017fsb,Banerjee:2022wht,Donnay:2021wrk,Donnay:2020guq,Stieberger:2018onx,Banerjee:2020kaa,Banerjee:2020zlg,Banerjee:2021cly,Gupta:2021cwo,Guevara:2021abz,Strominger:2021mtt,Himwich:2021dau,Melton:2022fsf,Adamo:2021lrv,Costello:2022wso,Costello:2022upu,Costello:2023vyy,Garner:2023izn,Stieberger:2022zyk,Stieberger:2023fju,Magnea:2021fvy,Gonzalez:2021dxw} has led to important insights into the study of scattering amplitudes rewritten as correlators of a conformal field theory on the 2D celstial sphere. This CFT is commonly known as the celestial conformal field theory (CCFT). The map from scattering amplitudes to the correlators of CCFT is done via the Mellin transformation. Usually scattering amplitudes are written in momentum basis. The job of Mellin transformation is to change the momentum eigenbasis into boost eigenbasis \cite{Pasterski:2016qvg,Pasterski:2017kqt,Banerjee:2018gce,Banerjee:2019prz}. The isomorphism between the global conformal group in 2D and the Lorentz group in 4D is at the heart of this basis change.

%CCFT is a potential candidate for the holographic dual of the 4D quantum gravity in asymptotically flat spacetime and it has received a great deal of attention lately, mainly due to its rich symmetry structure \cite{Strominger:2013lka,Strominger:2013jfa,He,Strominger:2014pwa,Campiglia:2016jdj,Kapec:2016jld, Kapec:2014opa, Banerjee:2020vnt,He:2017fsb,Banerjee:2022wht,Donnay:2021wrk,Donnay:2020guq,Stieberger:2018onx,Banerjee:2020kaa,Banerjee:2020zlg,Banerjee:2021cly,Gupta:2021cwo,Guevara:2021abz,Strominger:2021mtt,Himwich:2021dau,Melton:2022fsf,Ball:2021tmb,Adamo:2021lrv,Costello:2022wso,Costello:2022upu,Costello:2023vyy,Garner:2023izn,Stieberger:2022zyk,Stieberger:2023fju}. 

Operator Product Expansion (OPE) in CCFT  correspond to the collinear limit in the bulk and it plays a very important role in the study of the dual theory \cite{Fan:2019emx,Banerjee:2020zlg, Banerjee:2021cly, Guevara:2021abz, Himwich:2021dau, Costello:2022upu,Banerjee:2023zip,Ebert:2020nqf,Adamo:2022wjo,Ren:2023trv,Hu:2021lrx,Fan:2022vbz,Fan:2022kpp,Bhardwaj:2022anh,Krishna:2023ukw,Hu:2022bpa,Pate:2019lpp,Atanasov:2021cje, Banerjee:2023jne,Banerjee:2023rni,Ball:2023qim}. In a previous paper \cite{Banerjee:2023zip}, we have studied the $ w_{1+\infty} $ invariant OPEs in theories of gravity. Our analysis showed that there are an infinite number of theories on the celestial sphere which are $ w_{1+\infty} $ invariant. By deriving the OPE from graviton scattering amplitudes we have explicitly shown in \cite{Banerjee:2023jne} that the self dual gravity is one example of this infinite family.

In this paper we perform a similar analysis for gluons. In the case of gluons the infinite symmetry algebra is know as the $S$ algebra \cite{Guevara:2021abz,Strominger:2021mtt}. We write down all possible $S$ invariant OPE structures between two positive helicity outgoing gluons. We find that there is a (discrete) infinite number of such structures and presumably, each one of them corresponds to a $S$ invariant theory of gluons in the bulk. However, a more explicit Lagrangian description of these theories are not known to us. 

There is an important difference between the analyses of $ w_{1+\infty} $ and $ S $-invariant theories, which we want to point out. $ S $ algebra does not contain the Poincare generators. Therefore the consistent OPEs need not be Poincare invariant. However in this paper we make sure that all the OPEs are (conformal) Lorentz invariant and this plays an important role. This is along the line of \cite{Fan:2022vbz,Fan:2022kpp,Banerjee:2023rni,Casali:2022fro}.

We start with a brief review of the soft gluon symmetry algebra known as the $ S $ algebra in section \ref{sga}. In section \ref{gen_structure_OPE}, the general structure of the OPE between two positive helicity outgoing gluons on the celestial sphere has been discussed. We have argued, how the null states of the MHV-sector can be used to write down the general OPE. In section \ref{null_states}, we have written down the null states that appear at $ \mathcal{O}(z^0\bar z^0) $ of the gluon-gluon OPE in the MHV-sector. These are not the complete set of null states that the MHV sector has at $ \mathcal{O}(z^0\bar z^0) $. There are more of them. We talk about them later in section \ref{KZn}, where we have discussed the Knizhnik-Zamolodchikov (KZ)-type null states. Section \ref{OPE_organising} explicitly shows how to organise the OPE at every order. For simplicity we focus on the $ \mathcal{O}(z^0\bar z^0) $ terms in the OPE. We have also discussed the transformation properties of MHV-null states under the $ S $ algebra in this section, which are required to organise the OPE. Section \ref{OPEinv} shows the invariance of the $ \mathcal{O}(z^0\bar z^0) $ OPE under $ S $ algebra. In section \ref{inf}, we have argued, how an infinite number theories can exist on the celestial sphere. We conclude the paper with the discussion of the results found in this paper and some future directions in section \ref{dis}.

\section{The S algebra}
\label{sga}

We start by describing the soft symmetry algebra which follows from the universal singular terms in the OPE between two positive helicity outgoing gluons \cite{Guevara:2021abz,Strominger:2021mtt}. Let $ O^{a,+}_{\D}(z,\bar z) $ denote a positive helicity outgoing gluon conformal primary operator of dimension $ \D $ at the point $ (z,\bar z) $ on the celestial sphere. The universal singular terms in the OPE are given by
\be 
O^{a,+}_{\D_1}(z_1,\bar z_1) O^{b,+}_{\D_2}(z_2,\bar z_2) \sim \f{-i f^{abc}}{z_{12}} \sum_{n=0}^\infty B(\D_1-1+n,\D_2-1) \f{\bar z_{12}^n}{n!}\bar \pa^n O^{c,+}_{\D_1+\D_2-1}(z_2,\bar z_2)
\label{univ_OPE}
\ee
The interesting fact about these singular OPE coefficients is that it allows us to define an infinite tower of conformally soft \cite{Donnay:2018neh,Pate:2019mfs,Fan:2019emx,Nandan:2019jas,Adamo:2019ipt,Puhm:2019zbl,Guevara:2019ypd} gluons \cite{Guevara:2021abz,Strominger:2021mtt} by
\be
R^{k,a}(z,\bar z) := \lim_{\D\to k}(\D-k)O^{a,+}_{\D}(z,\bar z), \ k=1,0,-1,\cdots
\label{soft_gluon_curr}
\ee

Now it follows from the structure of the OPE \eqref{univ_OPE} that one can introduce the following truncated mode expansion
\be
R^{k,a}(z, \bar z) = \sum_{n=\f{k-1}{2}}^{\f{1-k}{2}} \f{R^{k,a}_n(z)}{\bar z^{n+\f{k-1}{2}}}
\label{soft_currents}
\ee
where the expansion coefficients $R^{k,a}_n(z)$ are the conserved holomorphic currents. For fixed values of $k$ and $a$ there are $(2-k)$ such currents and they transform in the $(2-k)$ dimensional representation of the $\overline{SL_2(R)}$. 

The holomorphic currents $R^{k,a}_n(z)$ can be further mode expanded as 
\be 
R^{k,a}_n(z) = \sum_{\a \in \mathbb{Z} - \f{k+1}{2}} \f{R^{k,a}_{\a,n}}{z^{\a+\f{k+1}{2}}}
\ee
The algebra of these modes can be obtained from the singular terms \eqref{univ_OPE} of the $RR$ OPE and is given by \cite{Guevara:2021abz},
\be
\begin{gathered}
\[R^{k,a}_{\a,n}, R^{l,b}_{\a',n'}\] = -i f^{abc} \f{\(\f{1-k}{2} - n + \f{1-l}{2} - n'\)!}{\(\f{1-k}{2} - n\)! \( \f{1-l}{2} - n'\)!}\f{\(\f{1-k}{2} + n + \f{1-l}{2} + n'\)!}{\(\f{1-k}{2} + n\)! \( \f{1-l}{2} + n'\)!}R^{k+l-1,c}_{\a+\a',n+n'}
\end{gathered} 
\label{comm_modes}
\ee
Now one can define the following generators \cite{Strominger:2021mtt}
\be
S^{q,a}_{\alpha,m} = \(q-m-1\)! \(q+m-1\)! R^{3-2q,a}_{\alpha,m} 
\ee
in terms of which the algebra \eqref{comm_modes} simplifies to 
\be
[S^{p,a}_{\alpha,m}, S^{q,b}_{\beta,n}] = -i f^{abc} S^{p+q-1,c}_{\alpha+\beta, m+n}
\ee
This infinite dimensional algebra of the conformally soft gluons, known as the $S$ algebra, plays a central role in this paper.\footnote{In this paper we write the OPE in terms of of the descendants of the $R$ algebra \eqref{comm_modes}. However, we will continue to refer to \eqref{comm_modes} as the $S$ algebra.}

In this paper we want to classify all possible OPEs between two positive helicity outgoing gluons which are invariant under the $S$ algebra. The strategy we adopt here is similar to that in the gravity case \cite{Banerjee:2023zip} but the details are very different. For example, in the gravity case the soft symmetry algebra which is isomorphic to the $w_{1+\infty}$ contains all four global space-time translations. But, this is not the case for the $S$ algebra and so there are $S$ invariant theories which are not space-time translationally invariant \cite{Fan:2022vbz,Fan:2022kpp,Banerjee:2023rni,Casali:2022fro}. However, we preserve Lorentz invariance because it translates into conformal invariance on the celestial sphere and the structure of the $S$ algebra depends on that.

\section{General structure of the OPE Between two positive helicity outgoing gluons}
\label{gen_structure_OPE}

 We can write the general structure of the OPE between two positive helicity gluons invariant under the $S$ algebra as 
\be 
\begin{gathered}
O^{a,+}_{\D_1}(z_1,\bar z_1) O^{b,+}_{\D_2}(z_2,\bar z_2) = \f{-i f^{abc}}{z_{12}} \sum_{n=0}^\infty B(\D_1-1+n,\D_2-1) \f{\bar z_{12}^n}{n!}\bar \pa^n O^{c,+}_{\D_1+\D_2-1}(z_2,\bar z_2) \\
+ \sum_{p,q=0}^\infty z_{12}^p \bar z_{12}^q \sum_{k=1}^{\tilde{n}_{p,q}}\tilde{C}_{p,q}^k(\D_1,\D_2) \tilde{O}^{ab}_{k,p,q} (\D_1,\D_2, z_2, \bar z_2) 
\end{gathered}
\label{gen_OPE}
\ee
where in the second line we have now added the $S$ algebra descendants of a positive helicity gluon. The sum over $ k $ could be finite or infinite depending on the theory. Our goal is to determine the descendants $ \tilde{O}^{ab}_{k,p,q} $ and the OPE coefficients $ \tilde{C}_{p,q}^k $ in a general $S$-invariant theory. 

In the gravity case \cite{Banerjee:2023zip} we found that any $w$-invariant OPE can be written in terms of the MHV OPE and the MHV null states. We have also checked by detailed calculation that this structure holds in the self-dual gravity theory \cite{Banerjee:2023jne} which is $w$ invariant. The same reasoning also goes through for the $S$ algebra and gluons. We summarize the argument below. 

Since the $S$ algebra is universal, i.e the \textit{same} \footnote{For example, this is not true in the conventional $2-D$ CFTs because different CFTs have different Virasoro central charges and so different conformal symmetry algebras.}  algebra holds in any $S$ invariant theory, it is reasonable to assume that there is a Master OPE which holds in \textit{all} $S$ invariant theories. Let us now consider the gluon-gluon OPE in the (tree-level) MHV sector of the pure YM theory. Since the MHV sector is $S$ invariant the Master OPE, when inserted in a MHV gluon scattering amplitude, should reproduce the known MHV sector OPE. Therefore one can write
\be
\text{Master OPE} = \text{MHV-sector OPE} + R
\ee
where $R$ should \textit{vanish} inside an MHV scattering amplitude. This is possible only if $R$ is a linear combination of MHV \textit{null states}. Now since the MHV-sector OPE already contains the universal singular terms \eqref{univ_OPE} of the gluon-gluon OPE, $R$ consists only of non-singular terms. So we can write,
\be 
\begin{gathered}
O^{a,+}_{\D_1}(z_1,\bar z_1) O^{b,+}_{\D_2}(z_2,\bar z_2)\big|_{\textnormal{Any Theory}} = O^{a,+}_{\D_1}(z_1,\bar z_1) O^{b,+}_{\D_2}(z_2,\bar z_2)\big|_{\textnormal{MHV}}  \\
+ \sum_{p,q=0}^\infty z_{12}^p \bar z_{12}^q \sum_{k=1}^{\tilde{n}_{p,q}}\tilde{C}_{p,q}^k(\D_1,\D_2) M^{ab}_{k,p,q} (\D_1,\D_2, z_2, \bar z_2) 
\end{gathered}
\label{gen_OPE_1}
\ee
where $ M^{ab}_{k,p,q} $ are the MHV null states. So when "Any Theory" is taken to be the MHV sector, $M^{ab}_{k,p,q}$ vanishes and we get back the MHV sector OPE by construction. 

We now describe the MHV null states which are of interest to us. In this paper we apply this general procedure to write down the OPE at $\mathcal{O}(z_{12}^0\bar z_{12}^0)$. 

\section{Null states in the MHV sector}
\label{null_states}
The general null state at order $ z_{12}^0 \bar z_{12}^0 $ is given by \footnote{These null states can be obtained by taking soft limits of the gluon-gluon MHV OPE \cite{Banerjee:2020vnt}. The relevant terms in the gluon-gluon OPE in the MHV sector which gives rise to these null states are given in \eqref{ope_mhv}.}
\be \label{psinull}
\begin{gathered}
\Psi^{ab}_j(\D) = R^{-j,a}_{\f{j-1}{2},\f{j+1}{2}}\mathcal{O}^{b,+}_{\D+j}-\f{(-1)^j j}{\G(j+2)}\f{\G(\D+j-1)}{\G(\D-2)} R^{1,a}_{-1,0}\mathcal{O}^{b,+}_{\D-1} \\
 - \f{(-1)^j}{\G(j+1)} \f{\G(\D+j-1)}{\G(\D-1)}R^{0,a}_{-\f{1}{2},\f{1}{2}} \mathcal{O}^{b,+}_{\D}
\end{gathered}
\ee
Here we have ignored the $ (p,q) $ index and have simply written $ M^{ab}_{k}  $ instead of $ M^{ab}_{k,0,0}  $ for the order $ z_{12}^0 \bar z_{12}^0 $ MHV null states.

Now it turns out that the following basis of null states
\be
\begin{gathered}
M_k^{ab}(\D) = \sum_{i=1}^k \f{1}{\G(k-i+1)}\f{\G(\D+k-1)}{\G(\D+i-1)}\Psi_i^{ab}(\D)
\label{M_null_basis}
\end{gathered} 
\ee
is more convenient because they transform nicely under the $ S $-algebra. We will discuss their transformation law in the next section. 

We conclude this section by defining the antisymmetric part of the null states $ M_k^{ab}(\D) $ as
\be\label{asm}
M^a_{k}(\D) = f^{abc}M_k^{bc}  ,
\ee

\section{Organizing the OPE at every order}
\label{OPE_organising}

Since the MHV sector is $S$ invariant the MHV null states must form a representation of the $S$ algebra. In other words every generator of the $S$ algebra must map any MHV null state to another MHV null state. Our analysis shows that this representation is reducible and different $S$ invariant theories correspond to different irreducible components of this representation. So our first job is to study the action of the $S$ algebra generators on the MHV null states. This is facilitated by the following observation. 
  
  We have discussed in Section \ref{sga} that the $ S $ algebra is generated by an infinite number of holomorphic soft currents $ \{R^{k,a}_p(z)\}  $ \footnote{Here $-p$ is the antiholomorphic weight of the current.} where $ k=1,0,-1,-2,\cdots. $ is the dimension $ \D $ of the soft operator and $ \f{k-1}{2} \leq p \leq \f{1-k}{2} $. For a fixed $ k $, the soft currents $  R^{k,a}_{\frac{1-k}{2}}, \cdots , R^{k,a}_{\f{k-1}{2}} $ transform in a $ (2-k) $ dimensional representation of the $ \overline{sl_2(R)} $ \footnote{Note that we are assuming the theory to be (conformal) Lorentz invariant.}. This can be seen from the following commutation relations:
\be
\begin{gathered}
\[H^0_{0,-1},R^{k,a}_{m,p}\] = \f{1}{2}(2p+k-3)
R^{k,a}_{m,p-1} \ \textnormal{for} \ p > \f{k-1}{2}, \qquad  \[H^0_{0,-1},R^{k,a}_{m,\f{k-1}{2}}\] = 0\\
\[H^{0}_{0,0},R^{k,a}_{m,p}\] = -2p \, R^{k,a}_{m,p}\\
\[H^0_{0,1},R^{k,a}_{m,p}\] = \f{1}{2}(2p-k+3) R^{k,a}_{m,p+1} \ \textnormal{for} \ p < -\f{k-1}{2}, \qquad  \[H^0_{0,1},R^{k,a}_{m,-\f{k-1}{2}}\] = 0
\end{gathered} 
\label{sl2wR}
\ee

Now let us consider the currents $ R^{1,a}_0, R^{0,a}_{\f{1}{2}}, R^{-1,a}_{1}, \cdots $ with the lowest $ \overline{sl_2(R)} $ weights. Starting from $ R^{1,a}_0 $ all the currents in this family can be obtained by applying the global subleading soft gluon operator $  R^{0,b}_{\f{1}{2},\f{1}{2}} $. This can be seen from the the following commutation relations
\be\label{sub}
\[ R^{0,a}_{\frac{1}{2},\frac{1}{2}}, R^{k,b}_{m,\frac{1-k}{2}}\] = - i f^{abc} (2-k) R^{k-1,c}_{m+\frac{1}{2},\frac{2-k}{2}}
\ee
Equations \eqref{sl2wR} and \eqref{sub} show that we can write any generator of the $S$ algebra as a sum of products of the generators $\( R^{1,a}_{n,0}, R^{0,a}_{\frac{1}{2},\frac{1}{2}}, H^0_{0,0}, H^{0}_{0,\pm 1}\)$. Therefore in order to study the action of the $S$ algebra generators on the MHV null states we just need to focus on these finite number of generators.

\begin{figure}[h!]
\begin{center}
\includegraphics[scale=0.3]{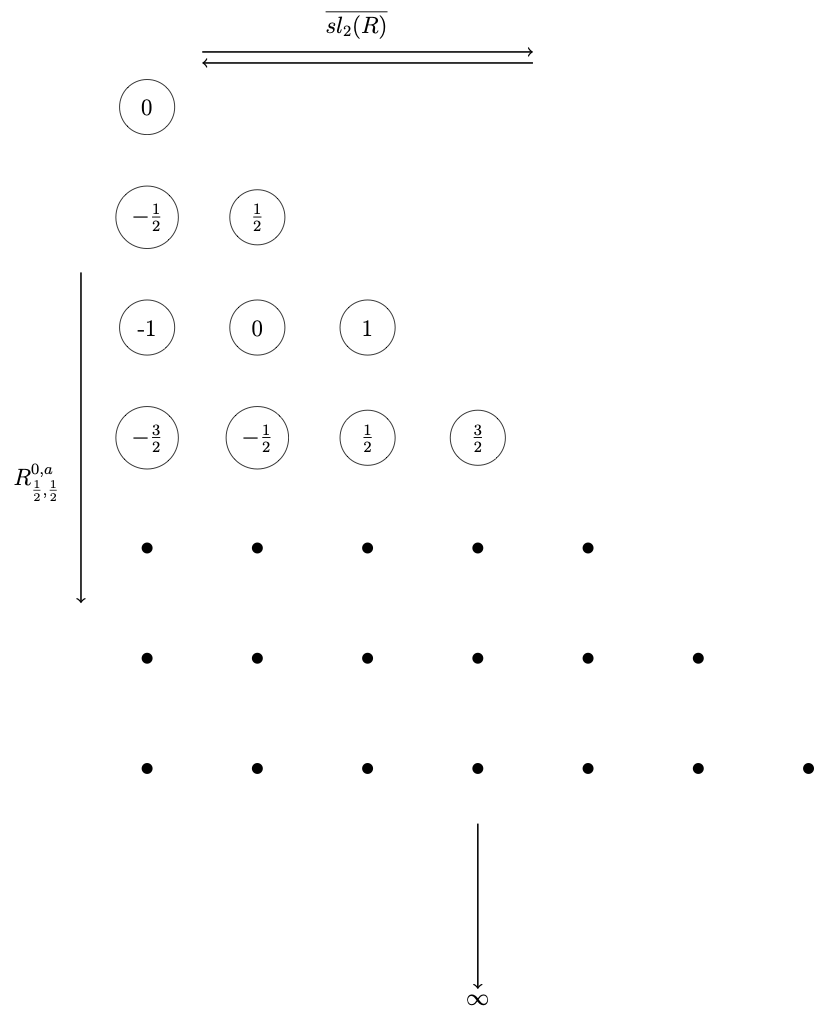}
\caption{The figure shows the soft gluon currents arranged in representations of $\overline{sl_2(R)}$. The $\overline{sl_2(R)}$ generators move the currents horizontally in both directions whereas the global subleading soft gluon symmetry generator $R^{0,a}_{\frac{1}{2},\frac{1}{2}}$ moves the currents vertically downward. }
\label{fig1}
\end{center}
\end{figure}

\subsection{Transformation properties of the null states under   $ \overline{sl_2(R)} $ algebra}

Using the action of different generators of $ \overline{sl_2(R)} $ algebra on the gluon primary operators and the commutation relations \eqref{sl2wR} it is easy to show that
\be
\begin{gathered}
H^0_{0,1} \Psi^{ab}_{k}(\D) = 0.
\end{gathered}
\label{eqn:54} 
\ee

Thus \eqref{M_null_basis} implies that
\be
\begin{gathered}
H^0_{0,1} M^{ab}_{k}(\D) = 0.
\end{gathered}
\label{eqn:55} 
\ee

Therefore the null states $M^{ab}_k$ are $\overline{sl_2(R)}$ primaries. 

\subsection{Transformation properties of the null states under the leading soft gluon current algebra}\label{rrl}

One can easily check that under the leading soft gluon current algebra, the null states \eqref{M_null_basis} transform as
\be
\begin{gathered}
R^{1,a}_{0,0} M_k^{bc}(\D) = - if^{abd}M_k^{dc}(\D) - if^{acd}M_k^{bd}(\D) \\
R^{1,a}_{n,0} M_k^{bc}(\D) = 0, \  n>0
\end{gathered} \label{trans}
\ee
Therefore the null states $M^{ab}_k$ are the leading soft gluon current algebra primaries.

\subsection{Transformation properties of the null states under subleading soft gluon operator $R^{0,a}_{\frac{1}{2},\frac{1}{2}}$}\label{trunc}

This is perhaps the most important transformation property because it mixes the null states $M^{ab}_{k}$ with different $k$ values. The action of  $ R^{0,a}_{\f{1}{2},\f{1}{2}} $ on $ M_{k}^{bc} (\D) $ is given by,
\be 
\begin{gathered}
R^{0,a}_{\f{1}{2},\f{1}{2}} M_k^{bc}(\D) = -(k+2)if^{abx}M_{k+1}^{xc}(\D-1) +(\D+k-2) \[if^{acx}M_k^{bx}(\D-1)+if^{abx}M_k^{xc}(\D-1)\]
\end{gathered}\label{tn}
\ee
We have used \eqref{act_sub_null_psi} to derive the above equation. 

Now let us consider the set of null states:
\be \label{set}
 M_k^{bc}(\D), \ k=1,2, \cdots,n .
 \ee
 From \eqref{tn} we can see that if we set
\be
\begin{gathered}
M^{ab}_{k+1}(\D) = 0, k\geq n \geq 0.
\end{gathered}
\label{cond:1} 
\ee
then the set \eqref{set} is closed under the action of $ R^{0,a}_{\f{1}{2},\f{1}{2}}$. Moreover it follows from \eqref{tn} that the infinite set of equations \eqref{cond:1} is also invariant under the action of $R^{0,a}_{\frac{1}{2},\frac{1}{2}}$ because the index $k$ mixes only with $k'\ge k$. Therefore the truncation \eqref{cond:1} is $S$ algebra invariant and we can get an $S$ invariant OPE if we keep only the finite set \eqref{set}. Let us emphasize that the integer $n$ is in no way restricted by the $S$ invariance. 

\section{$\mathcal{O}(z_{12}^0\bar z_{12}^0)$ OPE and its invariance under the $ S $ algebra}
\label{OPEinv}

Let us now consider the $ \mathcal{O}(1) $ terms in the OPE when we keep only the finite set of MHV null states \eqref{set}. In particular, we show that the $ \mathcal{O}(1) $ terms in the OPE with the following coefficients are $ S $-invariant:
\be
\boxed{
\begin{gathered}
\mathcal{O}^{a,+}_{\D_1}(z,\bar z) \mathcal{O}^{b,+}_{\D_2}(0,0)\big|_{\mathcal{O}(1)} = \mathcal{O}^{a,+}_{\D_1}(z,\bar z) \mathcal{O}^{b,+}_{\D_2}(0,0)\big|_{\textnormal{MHV OPE at  } \mathcal{O}(1)} \\
+ \sum_{k=1}^n B(\D_1+k,\D_2-1)M_k^{ab}(\D_1+\D_2) 
\end{gathered} 
\label{gen_OPE_coeff1}}
\ee

where $ \mathcal{O}^{a,+}_{\D_1}(z,\bar z) \mathcal{O}^{b,+}_{\D_2}(0,0)\big|_{\textnormal{MHV OPE at} \ \mathcal{O}(1)} $ is given by \cite{Banerjee:2020vnt,Ebert:2020nqf}
\be
\begin{gathered}
\mathcal{O}^{a,+}_{\D_1}(z,\bar z)\mathcal{O}^{b,+}_{\D_2}(0,0)\big|_{\textnormal{MHV OPE at} \ \mathcal{O}(1)} = B(\D_1-1,\D_2-1)\left[ \D_1 \, R^{1,a}_{-1,0}\mathcal{O}^{b,+}_{\D_1+\D_2-1}(0,0) \right.\\
\left.  + \f{\D_1-1}{\D_1+\D_2-2}R^{0,a}_{-\f{1}{2},\f{1}{2}}\mathcal{O}^{b,+}_{\D_1+\D_2}(0,0)\right]
\label{ope_mhv}
\end{gathered} 
\ee

Let us first apply $ R^{0,a}_{\f{1}{2},\f{1}{2}} $ on the OPE \eqref{gen_OPE_coeff1}. After some straightforward algebra we get,
\be
\begin{gathered}
R^{0,x}_{\f{1}{2},\f{1}{2}} \( \mathcal{O}^{a,+}_{\D_1}(z,\bar z) \mathcal{O}^{b,+}_{\D_2}(0,0)\big|_{\mathcal{O}(1)} \) - R^{0,x}_{\f{1}{2},\f{1}{2}} \[ \mathcal{O}^{a,+}_{\D_1}(z,\bar z) \mathcal{O}^{b,+}_{\D_2}(0,0)\big|_{\textnormal{MHV OPE at  } \mathcal{O}(1)} \right. \\
\left. + \sum_{k=1}^n B(\D_1+k,\D_2-1)M_k^{ab}(\D_1+\D_2) \]  = if^{xay} (n+2) B(\D_1+n,\D_2-1) M^{yb}_{n+1}(\D_1+\D_2-1)
\end{gathered} \label{r0}
\ee

Now, we have argued in the previous section that if the $ \mathcal{O}(1) $ OPE of a $S$ invariant theory truncates at $ k=n $, then $ M^{ab}_{n+1}(\D) $ will be a null state of that theory. Thus, we can set the RHS of \eqref{r0} to 0 and hence \eqref{gen_OPE_coeff1} is invariant under the action of $ R^{0,a}_{\f{1}{2},\f{1}{2}} $. Using \eqref{trans}, one can also verify that, \eqref{gen_OPE_coeff1} is invariant under the actions of $ R^{1,a}_{n,0}$.

In \cite{Banerjee:2020vnt}, it was shown that the OPE in the MHV-sector is invariant under the action of $ H^0_{0,1} $. We can also see from \eqref{eqn:54} that the null states $ M_k^{ab}(\D) $ are annihilated by $ H^0_{0,1} $. Thus, the OPE \eqref{gen_OPE_coeff1} is also invariant under $ H^0_{0,1} $. Hence we conclude that the truncated OPE \eqref{gen_OPE_coeff1} is invariant under the $ S $ algebra.

\section{Infinite family of $ S $ invariant theories}
\label{inf}

In section \eqref{trunc} we have shown that the following set of equations
\be
\begin{gathered}
M^{ab}_{k+1} = 0, k\geq n \geq 0.
\end{gathered}
\label{cond:11}
\ee
are $ S $ invariant. Thus at $ \mathcal{O}(z^0 \bar z^0) $ we can truncate the OPE \eqref{gen_OPE_1} at an arbitrary $ n $ in an $ S $-invariant way. That is to say $ S $-invariance does not fix the value of the integer $ n $. Hence we can get a discrete infinite family of $ S $-invariant OPEs for different choices of the integer $ n $. Each of these consistent OPEs correspond to a $S$ invariant theory of gluons. But, at present we do not know the Lagrangian description of these theories except perhaps the self-dual Yang-Mills theory. 

\section{Knizhnik-Zamolodchikov type null states}
\label{KZn}

Knizhnik-Zamolodchikov (KZ) type null states contain descendants of the holomorphic translation generator $L_{-1}$ on the celestial sphere. They can be obtained algebraically by determining the relevant primary descendant but in our case we can bypass this tedious procedure if we use the OPE commutativity
\be
\begin{gathered}
\mathcal{O}^{a,+}_{\D_1}(z_1,\bar z_1) \mathcal{O}^{b,+}_{\D_2}(z_2,\bar z_2) = \mathcal{O}^{b,+}_{\D_2}(z_2,\bar z_2) \mathcal{O}^{a,+}_{\D_1}(z_1,\bar z_1)
\label{comm_OPE}
\end{gathered} 
\ee
The reason behind this is that the $\mathcal{O}(z_{12}^0\bar z_{12}^0)$ terms of the OPE, as written in \eqref{gen_OPE_coeff1}, are not manifestly symmetric under the exchange \eqref{comm_OPE}. Therefore OPE commutativity imposes non-trivial constraints on the OPE coefficients and one such constraint is essentially the KZ equation. The process can be further simplified if make the operator $\mathcal{O}^{b,+}_{\D_2}(z_2,\bar z_2)$ leading soft by taking the limit $ \D_2 \to 1 $. Now a straightforward calculation gives the KZ type null state

\begin{equation}
\boxed{
	\label{eq:ten6}
	\begin{split}
K^{a}(\D)	= \xi^a(\D) - i\sum_{k=1}^{n} M^{a}_{k}(\Delta+1)=0.
	\end{split}}
\end{equation}
where
\be\label{kzmhv}
\xi^a(\D) = C_{A}L_{-1}\mathcal{O}^{a,+}_{\Delta}-(\Delta+1)R^{1,b}_{-1,0}R^{1,b}_{0,0}\mathcal{O}^{a,+}_{\Delta}-R^{0,b}_{-\frac{1}{2},\frac{1}{2}}R^{1,b}_{0,0}\mathcal{O}^{a,+}_{\Delta+1}
\ee
is the KZ type null state in the MHV-sector \cite{Banerjee:2020vnt} and $ M^a_k(\D) $ is the antisymmetric part of the null state $ M^{ab}_k(\D) $ defined in \eqref{asm}. We have also used the identity $ f^{abx} f^{aby} = C_A \d^{xy} $ in deriving the KZ type null state equation \eqref{eq:ten6}.

Another null state equation involving the descendant $ L_{-1}\mathcal{O}^{a,+}_{\Delta} $ can be obtained from \eqref{comm_OPE} in a similar way by taking the subleading conformal soft limit $ \D_2 \to 0 $. It is given by
\begin{equation}
	\label{eq:ten7}
	\begin{split}
		(\D-1) \xi^a(\D) - \sum_{k=1}^{n} (\D+k) M^{a}_{k}(\Delta+1)=0.
	\end{split}
\end{equation}

Now multiplying equation \eqref{eq:ten6} by $ (\D-1) $ and then subtracting it from \eqref{eq:ten7} we get the following (current algebra) null state 
\be
\label{mn}
\chi^{1,a}_n(\D) = \sum_{k=1}^n (k+1) M_{k}^{a}(\D) = 0 
\ee
One can continue this procedure and get other current algebra null states by taking conformal soft limits $\D_2\rightarrow k, \ k\le -1$. We can denote them by $ \{\chi^1_n(\D), \chi^2_n(\D), \cdots \} $. However, it can be shown that after a finite iteration this procedure stops due to the truncation \eqref{cond:1}.

\subsection{S invariance of the KZ type null state}

In this section we show that the KZ type null state \eqref{eq:ten6} is $ S $-invariant. 

First of all, the states $ M_k^{ab}(\D) $, and as a result $ M^a_k(\D)= f^{abc}M^{bc}_k(\D) $, are annihilated by $ H^0_{0,1} $. Therefore the state $ K^a(\D) $ is a primary of $ \overline{sl_2(R)} $ because the KZ type null state \eqref{kzmhv} in the MHV-sector is annihilated \cite{Banerjee:2020vnt} by $ H^0_{0,1} $.

Similarly, one can show after some algebra that the following relation holds
\be
\begin{gathered}
R^{0,c}_{\f{1}{2},\f{1}{2}} \ K^{a}(\D) = (\D-2) if^{cax} K^{x}(\D-1) - (n+2) f^{cbx} f^{aby} M^{xy}_{n+1}(\D) + f^{cax} \chi^{1,x}_n(\D) \\
+ f^{cby}f^{bax} \[ \sum_{k=1}^n (k+1)M_k^{yx}(\D) + 2 E^{yx}(\D) \] + f^{cab}f^{byx}  E^{yx}(\D)
\end{gathered} 
\ee
where \be\label{eeq} E^{yx}(\D) = (\D-2) R^{1,y}_{-1,0} \mathcal{O}^{x,+}_{\D-1} + R^{0,y}_{-\f{1}{2},\f{1}{2}}\mathcal{O}^{x,+}_{\D} \ee

Now we know that, in a theory in which the $\mathcal{O}(z_{12}^0\bar z_{12}^0)$ OPE truncates at  $ k=n $, i.e, \eqref{cond:1} holds, both $ M_{n+1}^{bc}(\D) $ and $ \chi^{1,a}_n(\D) $ are null states. Thus we can set them to 0 and get,
\be
\begin{gathered}
R^{0,c}_{\f{1}{2},\f{1}{2}} \ K^{a}(\D) = (\D-2) if^{cax} K^{x}(\D-1) + f^{cby}f^{bax} \[ \sum_{k=1}^n (k+1)M_k^{yx}(\D) + 2 E^{yx}(\D) \] + f^{cab}f^{byx}  E^{yx}(\D) 
\end{gathered} \label{kz1}
\ee
We show in Appendix \eqref{KZS} that the second and the third terms on the RHS of \eqref{kz1} are actually zero. Taking this into account we get
\be
\begin{gathered}
R^{0,c}_{\f{1}{2},\f{1}{2}} \ K^{a}(\D) = (\D-2) if^{cax} K^{x}(\D-1) 
\end{gathered} \label{kz2}
\ee
Thus we see that $ R^{0,c}_{\f{1}{2},\f{1}{2}} $ maps the KZ type null state $K^a(\D)$ to linear combination of other null states in the theory. Hence, the null state equation
\be 
K^{a}(\D) = 0
\ee
is $S$ invariant. 

%\section{Discussion}
\label{dis}

\section{Acknowledgement}
SB would like to thank the participants of the Kickoff Workshop for the Simons Collaboration on Celestial Holography for helpful comments. The work of SB is partially supported by the  Swarnajayanti Fellowship (File No- SB/SJF/2021-22/14) of the Department of Science and Technology and SERB, India. The work of PP is supported by an IOE endowed Postdoctoral position at IISc, Bengaluru, India. SP is supported by the INSA Senior scientist position at NISER, Bhubaneswar through the Grant number INSA/SP/SS/2023.

\appendix

\section{$ S $ algebra primaries}

In this Appendix, we write down the conditions on the primary operators that follow from the OPE between two positive helicity outgoing gluon primaries \eqref{univ_OPE}. They are obtained by taking different soft limits in \eqref{univ_OPE} and comparing both the sides of the OPE:
\be \label{eq:A1}
\begin{split}
R^{k,a}_{p-\f{k+1}{2}, - q - \f{k-1}{2}}\mathcal{O}^{b,+}_{\D}(0,0) &= 0, p\geq 2 \\
R^{k,a}_{\f{1-k}{2}, - q - \f{k-1}{2}}\mathcal{O}^{b,+}_{\D}(0,0) &= - i f^{abc}\f{(-1)^{k+q+1}}{\G(-k-q+2)}\f{\G(\D-1)}{\G(\D+q+k-2)}\f{\bar \pa^q}{q!} \mathcal{O}^{c,+}_{\D+k-1}(0,0)
\end{split} 
\ee

where $ 0\leq q \leq 1-k, \ k=1,0,-1,\cdots $. These conditions have been used in writing down the transformation properties of the MHV null states. For more details of how to obtain these conditions one can check Appendix F of \cite{Banerjee:2023jne}. The analyses there was done for $ w_{1+\infty} $ primaries, but the methodology is same for $ S $ algebra also.

\section{Transformation properties of the $ \Psi_j^{bc} $-null states under the leading soft gluon operator $ R^{1,a}_{0,0} $ and the subleading soft gluon operator $ R^{0,a}_{\f{1}{2},\f{1}{2}} $}
\label{primary_check}
Using \eqref{psinull}, \eqref{comm_modes} and \eqref{eq:A1}, one can show that,
\be
\begin{split}
R^{1,a}_{0,0} \Psi^{bc}_j(\D) &= -if^{abx} \Psi^{xc}_j(\D) - if^{acx} \Psi^{bx}_j(\D)  \\
R^{0,a}_{\f{1}{2},\f{1}{2}}\Psi^{bc}_j(\D) &=  - (j+2) \, if^{a b x}\Psi^{xc}_{j+1}(\D-1)+(\D+j-2) \, if^{a c x}\Psi^{bx}_{j}(\D-1)\\
& + 2\f{(-1)^j}{\G(j+1)}\f{\G(\D+j-1)}{\G(\D-1)}if^{a b x} \Psi^{xc}_1(\D-1)
\label{act_sub_null_psi}
\end{split} 
\ee

These equations have been used in sections \ref{rrl} and \ref{trunc}.

\section{Proof that the KZ-type null states are closed under the action of $ R^{0,a}_{\f{1}{2},\f{1}{2}} $}\label{KZS}

We write the second and third term in \eqref{kz1} as,
\be
\begin{gathered}
\Sigma^{ca}(\D) = f^{cby}f^{bax}\[\sum_{k=1}^n(k+1) M_k^{yx}(\D)+2E^{yx}(\D)\] + f^{cab}f^{byx} E^{yx}\(\D\)
\end{gathered} \label{eqc1}
\ee

The above equation can be decomposed into symmetric and antisymmetric part in the following way,
\be
\Sigma^{ca}(\D) = \Sigma_A^{ca}(\D) + \Sigma_S^{ca}(\D) 
\ee

where
\be
\begin{split}
\Sigma_A^{ca}(\D) &= \f{1}{2}\[\Sigma^{ca}(\D) - \Sigma^{ac}(\D)\] \\
\Sigma_S^{ca}(\D) &= \f{1}{2}\[\Sigma^{ca}(\D) + \Sigma^{ac}(\D)\] \label{sp}
\end{split} 
\ee

Now, using the Jacobi identity
\be 
f^{acb}f^{bxy}+f^{xab}f^{bcy}+f^{xcb}f^{aby}=0
\ee
one can show that
\be
\Sigma_A^{ca}(\D) = -\f{1}{2} f^{cab}\chi^{1,b}_n(\D) = 0
\ee

We now simplify the symmetric part \eqref{sp} and get,
\be
\Sigma_S^{ca}(\D) = \f{1}{2}f^{cby}f^{bax} \[ \sum_{k=1}^n (k+1) \(M_k^{xy} (\D) + M_k^{yx}(\D)\) + 2 \(E^{xy} (\D) + E^{yx}(\D)\) \] 
\ee

The leading and subleading soft limits of \eqref{comm_OPE} and some straightforward algebra then gives,
\be
\sum_{k=1}^n (k+1) \(M_k^{xy} (\D) + M_k^{yx}(\D)\) + 2 \(E^{xy} (\D) + E^{yx}(\D)\)  = 0. 
\ee

Hence we conclude that,
\be
\Sigma^{ca}(\D) = 0.
\ee

\end{document}